# CRITICAL CURRENT ANISOTROPY OF PRACTICAL SUPERCONDUCTORS: ANALYSIS METHODS AND APPLICATION CASES

V.V. Guryev, I.V. Kulikov, S.V. Shavkin
NRC «Kurchatov Institute», Moscow, Russia

The analysis of the critical current anisotropy holds significant importance in the process of enhancing the efficiency of devices that rely on superconductors. The article delivers a concise yet critical evaluation of contemporary techniques utilized for analyzing the critical current angular dependence of practical superconductors, with a specific focus on second-generation high-temperature superconducting tapes (coated conductors). These techniques are based either on the scaling model, the vortex path model, or the anisotropic pinning model. The paper presents findings from an empirical investigation into the critical current angular dependencies of coated conductors with different chemical compositions. Several distinctive features are highlighted, including the intricate impact of substituting the rare-earth element in the HTS composition on the pinning landscape, the asymmetry of the peaks, and the dependence of the critical current value on the Lorentz force direction at fixed magnetic field direction. Then, the described analysis techniques are applied to the experimental dataset to derive meaningful insights. A refined version of the anisotropic pinning model is introduced to accurately depict certain observed phenomena. The accuracy of approximations produced by various models is evaluated using the coefficient of determination, adjusted to account for the number of variables used for fitting. The fundamental disparity in the interpretation of angular dependencies when employing different models is underscored. The lack of a universal methodology that can comprehensively explain all features while establishing a connection with the defective structure of the HTS material is stated. This signifies a critical gap in the existing understanding of the behavior of superconductors under varying conditions and highlights the need for further research and development in this area of study.

## Introduction

The critical current ($I_c$) of practical superconductors, and in particular high-temperature tapes of the second generation (HTS-2 or coated conductors), depends on both the magnitude and the direction of the magnetic field. In the technology pertaining to coated conductors, a high-temperature superconductor layer (RE)$Ba_2Cu_3O_{7-x}$ (REBCO, where RE denotes a rare earth element) is epitaxially deposited onto textured buffer layers that cover a flexible metallic substrate. The anisotropy must be taken into account when designing devices. For example, when assembling a solenoid from double pan cakes, different turns in the outer pancake are in significantly different conditions: the inner turns are subjected to a magnetic field that is nearly aligned with the plane of the tape, which typically corresponds to the maximum critical current, whereas in the outer turns, the magnetic field lines are tilted, resulting in a reduction of the critical current, even though the intensity of the magnetic field in this region is relatively lower (Fig. 1).

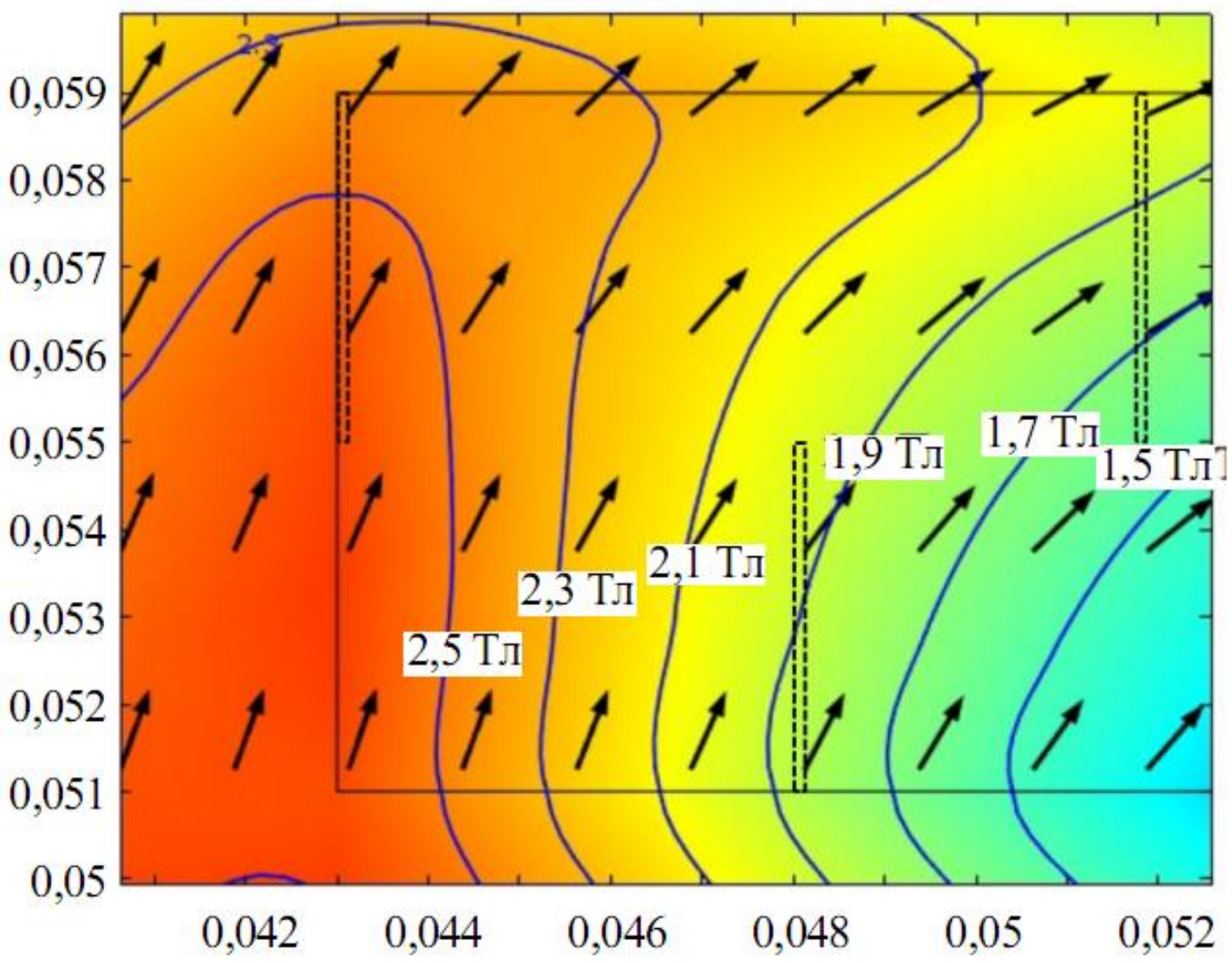

Fig. 1. The computed distribution of the magnetic field around the inner edge of the outer double pancake of the HTS solenoid [1, 2]: the solenoid axis is on the left; ↗ — local direction of magnetic field; — — lines of constant magnetic field color changes from colder to hotter in accordance with the increase in magnetic field; ---- — denotes the coated conductor tapes incorporated within the winding of the double pancake. The current within the tapes traverses perpendicular to the plane of the illustration.

Note that in the entire solenoid including both the internal and external pancakes, as well as the internal and external turns of each pancake, the magnetic field is oriented orthogonally to the current. Therefore, the angular dependence of the critical current $I_c(\theta)$, where $\theta$ is the angle between the tape normal and the direction of the magnetic field, while preserving the orthogonality of the transport current and magnetic field, holds paramount significance in the design of solenoids. The experimental geometry, whereby the orientations of the current and external magnetic field remain perpendicular as the angle θ varies, is frequently referred to as the configuration of maximum Lorentz force.

The angular dependence $I_c(\theta)$ of coated conductors is influenced by the synergistic impact of intrinsic anisotropy, caused by the structure of the REBCO crystal unit cell, and the external anisotropy, dictated by the pinning landscape attributes of specific superconductors. Distinguishing the contributions of these two factors is an extremely complex problem [3]. Currently, the angular dependences $I_c(\theta)$ are considered as an attribute of a coated conductor, determined experimentally after tape production. A comprehensive database has been established for commercially available tapes to facilitate the selection of a tape for a specific application [4]. However, the holy grail of practical superconductivity is the development of a technology that would allow fine-tuning the current-carrying capacity of a superconductor to suit a customer's order even before actual production. The key point on the way to achieving this goal is the development of methods for analyzing the angular dependences $I_c(\theta)$ to establish its relationship with the structure.

Below, we give a brief critical review of the known methods for analyzing the critical current angular dependences. For a more comprehensive understanding of the historical context surrounding some of these methods, we recommend that readers consult the review provided in [3]. Then we give a description of the experimental techniques employed; describe the samples, and present experimental angular dependences of coated conductors with varying chemical compositions. These experimental angular dependences are analyzed using the described methods. The findings of the work done are summarized in the conclusion.

## Methods of critical current angular dependencies analysis

A diverse array of approximating functions for $I_c(\theta)$ [5-7] and numerical description techniques [8-10] have been proposed to describe the critical current angular dependencies, which may facilitate engineering calculations. However, the fitting coefficients derived from these approximations frequently lack physical meaning and, therefore, cannot be correlated with the material's structural attributes. In our opinion, such approaches do not contribute to understanding the relationship between structure and current-carrying capacity and will not be considered in this paper.

**Scaling method.** The scaling method that accounts for the anisotropy inherent in the effective mass of charge carriers, also known as the Blatter scaling method, represents the historically first and still the most prevailing method for analyzing the critical current angular dependencies. This approach assumes that the angular dependence $I_c(\theta)$ is primarily dictated by the intrinsic anisotropy of the material, rather than the configuration of the pinning center system. As a result, the angular dependence $I_c(\theta)$ obeys the same functional dependence as the thermodynamic critical parameters, such as the upper and lower critical fields:

$$f(\theta) \sim (\cos^2\theta + \frac{\sin^2\theta}{\gamma^2})^{-1/2} \tag{1}$$

where $\gamma = \sqrt{m_c/m_{ab}}$ is the anisotropy parameter, $m_c$ and $m_{ab}$ - effective masses of charge carriers in c - direction and in ab-plane of unit cell of HTS respectively [11]. The parameter γ can, in principle, be computed based on information about the structure of the crystal cell. In the case of coated conductors based on YBCO, this parameter takes a value in the range from 5 to 7. Furthermore, the value of γ is independent of temperature and magnetic field [12]. Another prediction of this model is that the critical current $I_c$ depends on the value of $H$ and the orientation angle $\theta$ through a combined single variable: $I_c(H,\theta) = I_c(\widetilde{H})$, where $\widetilde{H} = H/f(\theta)$ [13 With an appropriately selected fitting parameter γ, the angular dependencies measured at varying magnetic field strengths and temperatures should align into a single so-called scaling curve when reconstructed in the coordinate $I_c - \widetilde{H}$. Any potential deviations from this scaling curve are interpreted as contributions from correlated pinning centers. [14].

Despite the widespread use of this approach, there are serious doubts about its theoretical foundations [3]. The theoretical framework of weak collective pinning, which serves as the foundation for this model, is applicable solely for minimal deformations of the vortex lattice and, accordingly, small currents. For up-to-date coated conductors, the critical currents attain approximately 10% of the Cooper pair decoupling current, resulting in such pronounced deformation of the vortex system that it is more accurately described as an amorphous vortex medium rather than a lattice. In addition, the parameter γ often exhibits values that substantially deviate from the anticipated range of 5-7, thereby complicating its interpretation.

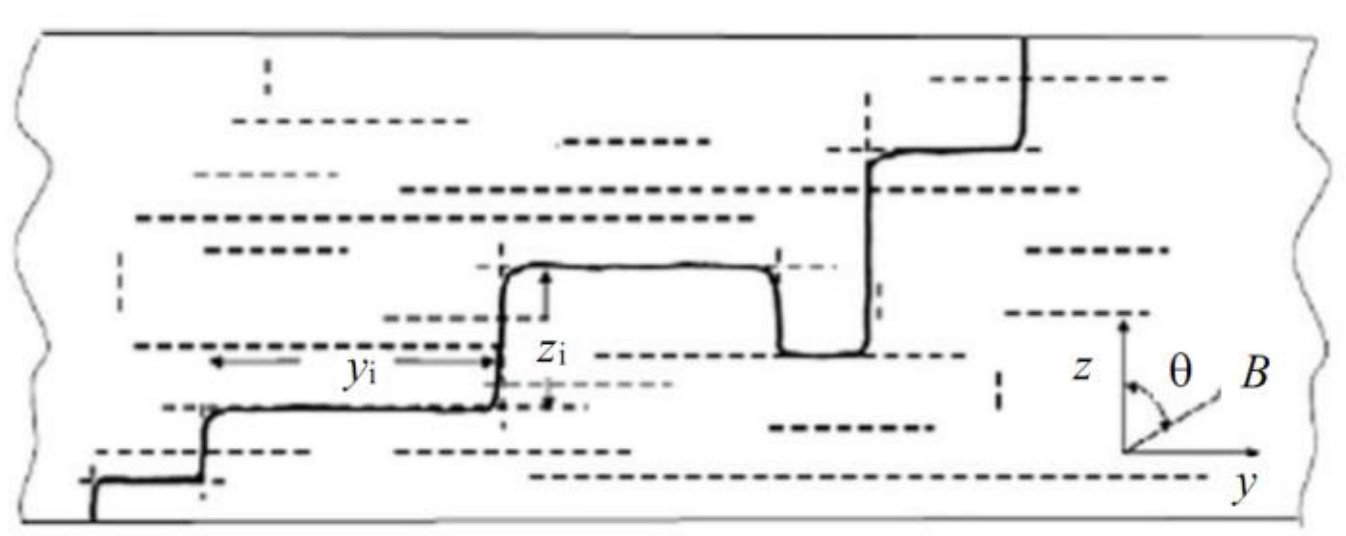

Fig. 2. Possible position of a pinned vortex — the so-called “vortex path”: ---- —defects (pinning centers) in the material [18]. Pinning occurs in the y—z plane so that, on average, the vortex is oriented at a macroscopic angle θ to the z axis.

The model claims that anisotropy is mainly determined by thermodynamics and, thus yielding a rather pessimistic forecast for the efficiency of manipulating the pinning center system to control anisotropy. Conversely, experimental data indicate that through the strategic control of the pinning landscape, it is possible to significantly modify the anisotropy of REBCO tapes [15] up to obtaining the inverse anisotropy [16] or its complete elimination [17]. Therefore, at least in the case of not too strong internal anisotropy (which is true for REBCO), it appears warranted to revise the foundational assumptions: to prioritize the interaction of vortices with pinning centers as the principal mechanism, while considering the impact of internal anisotropy as a minor modification.

**Vortex path model.** The theoretical framework underlying the vortex path model is described as follows [3, 18, 19]. It is assumed that the pinning landscape possesses a disordered and directionally correlated nature, as depicted in Fig. 2. The "vortex path" illustrated in Fig. 2 represents a possible state of a pinned vortex. On average, the orientation of the vortex aligns with the external magnetic field at a macroscopic angle $\theta$. It is postulated that the bulk pinning force and, hence, the current density $J_c$ are proportional to the density of such paths.

The vortex path is conceptualized as a directed stochastic walk. The probability density of the vortex path is defined as $p(\theta)$, while $y = n\lambda$ denotes the cumulative result of $n$ steps of average length $\lambda$ in the $y$ direction. It is assumed the absence of directional bias for steps in the z direction and that the probability of finding a particular value $z = \sum_{i=1}^{n} z_i$: $p(z) = (1/2\pi n\sigma^2)^{1/2}\exp(-z^2/2n\sigma^2)$ is a Gaussian distribution. Since $z/y = \tan(\theta)$, by transforming the distribution variable from $J_c(z/y) \propto p(z)$ to $J_c(\theta) \propto p(\theta)$ one derives

$$J_c(\theta) = \frac{J_0}{\sqrt{2\pi}\Gamma sin^2\theta} \times \exp(-\frac{1}{2\Gamma^2 tan^2\theta}) \quad (2)$$

where $\Gamma=\sigma/\sqrt{n}\lambda$; $J_0$ represents the proportionality constant. This relationship is referred to as the angular Gaussian. If, instead of assuming convergence to a Gaussian for p(z), the Lorentz distribution $p(z) = (1/\pi)\gamma/(\gamma^2 + z^2)$ is chosen, then the result will be:

$$J_c(\theta) = \frac{1}{\pi}\frac{J_0\Gamma}{cos^2\theta + \Gamma^2 sin^2\theta} \quad (3)$$

In this case, the fitting coefficient $\Gamma = \gamma/\lambda$. Formula (3) is designated as the angular Lorentzian.

The vortex path model presupposes the representation of the experimental curve $I_c(\theta)$ through a combination of a Gaussians and/or a Lorentzians, and/or a linear combination of them with different fitting parameters. In this manner, it is possible to achieve a robust approximation of $I_c(\theta)$ across a wide range of technical superconductors, irrespective of the presence or absence of internal anisotropy [3, 20, 21]. It is hypothesized that each Gaussian and Lorentzian corresponds to its distinct subsystem of pinning centers (PC). This assumption is the strongest aspect of the approach, since it allows identifying subsystems of PC and enables inferences regarding the evolution of their efficacy under varying external conditions. It is noteworthy that the each Gaussian or Lorentzian does not necessary correspond to a single peak in the angular dependency of Ic(θ), but always defined by the single PC subsystem. Thus, the asymmetric peaks of $I_c(\theta)$ in this model is characterized by a specific collection of Gaussians or Lorentzians with close positions of the average, and therefore by several PC subsystems. Conversely, if the value of the parameter Γ exceeds unity, the angular Gaussian (2) has two peaks in the range from 0 to π, and thus one CP subsystem is responsible for both of these peaks [19].

It should be noted that, according to experimental observations, the inclusion of an additional subsystem of pinning centers does not always lead to an additive effect on the critical current angular dependence [22]. Consequently, the model's assumption that the influence of various subsystems of pinning centers can be delineated through a linear combination of Gaussians and/or Lorentzians necessitates additional justification. The complexity of the situation is exacerbated by the lack of strict criteria for determining the best approximation, which leads to a certain amount of uncertainty due to subjective decisions when processing angular dependences [23]. This leads to difficulties in determining the number of PC subsystems.

The theoretical basis of this method is also questionable. A high density of vortex paths will result in the constancy of total energy during minimal vortex displacements, consequently leading to a reduction rather than an augmentation of the volume pinning force. In addition, it is well known that the single-vortex approximation shown in Fig. 2 inadequately represents the dynamics of a vortex matter, even at a qualitative level [24]. In a practical context, the curvatures depicted in Fig. 2 will be obstructed by adjacent vortices especially as the field increases and the distance between vortices diminishes.

**Anisotropic pinning model.** An approach that allows both to bypass the limitations of the single-vortex approximation and to avoid the summation problem was proposed for the first time in [25]. This model does not focus on individual pinning mechanisms; rather, it addresses the pinning of a sufficiently large ensemble of vortices situated within a potential well, which is specified by the cooperative influence of all pinning centers present in the volume of this ensemble. In the absence of a transport current, the vortex ensemble occupies the most favorable position at the bottom of the cooperative potential well, whereby the minimum specific energy of the magnetic flux at rest does not imply that each vortex captured by the nearest pinning centers is at the bottom of its own potential well. Under the action of the transport current, the vortex ensemble rises along the gradient of the cooperative potential well in the orientation dictated by the Lorentz force. If the Lorentz force exceeds the maximum steepness of the slope of the cooperative well, all vortex matter begins to move, resulting in the generation of an electric field and energy dissipation. Thus, the volume pinning force can be expressed as

$$F_p = -\max\left(\frac{\delta U}{\delta L}\right) = -\vec{e}_l \frac{U_0(\vec{B})}{L_0(\vec{j}, B)} \tag{4}$$

where $U$ is the depth of the cooperative potential well; el is the unit vector in the direction of the Lorentz force; $L_0 = U_0/\max(\delta U/\delta L)$ is the effective size of the cooperative potential well; $U_0$ is the effective depth of the cooperative potential well; j is the current density vector.

The model assumes that the effective depth of the cooperative potential well $U_0(\vec{B})$) depends only on the magnitude and direction of the magnetic induction, but not on the Lorentz force. The width of the cooperative potential well $L_0(\vec{j}, B)$, on the contrary, is determined by the absolute value of the magnetic induction (the density of vortices) and the direction of the Lorentz force. Thus, the model accounts for anisotropy not only in relation to the direction of the magnetic field but also in relation to the orientation of the Lorentz force, which is a strong point of the model. The effective width of the cooperative potential well $L_0$ is determined by the minimum distance between two energetically equivalent positions of the vortex ensemble. As a rough estimate of this distance, one can take the minimum of two values: the average intervortex distance or the mean distance between adjacent (in the direction of possible motion) pinning centers.

To acquire a comprehensive understanding of the current-carrying capacity, it is necessary to reconstruct the orientation, field and temperature dependences $U_0$ and $L_0$ of the so-called energy and dimensional bodies. Based on a large number of complementary experiments [24-29], it was shown that for a superconducting Nb-Ti tape, the angular dependences $U_0$ and $L_0$ are described by ellipsoids:

$$\left(\frac{\cos(\alpha)}{U_x}\right)^2 + \left(\frac{\cos(\beta)}{U_y}\right)^2 + \left(\frac{\cos(\gamma)}{U_z}\right)^2 = \frac{1}{U_0^2} \tag{5a}$$

where cos(α), cos(β), cos(γ) are the direction cosines of the induction vector. Similarly for the dimensional body:

$$(\frac{\cos(\alpha')}{L_x})^2 + (\frac{\cos(\beta')}{L_y})^2 + (\frac{\cos(\gamma')}{L_z})^2 = \frac{1}{L^2} \quad (5b)$$

where cos(α′), cos(β′), cos(γ′) are the direction cosines of the Lorentz force vector.

To determine the critical value of the pinning force under given conditions, it is necessary to divide the value of the energy ellipsoid radius (in the direction of the induction vector) by the value of the dimensional ellipsoid radius (in the direction of the Lorentz force).

For the experimental geometry corresponding to the configuration of the maximum Lorentz force, this model leads to the angular dependence of the critical current [28]:

$$J_c(\theta) = J_c(90^\circ)\sqrt{\frac{(k^L * \cos(\theta))^2 + \sin(\theta)^2}{(k^U * \cos(\theta))^2 + \sin(\theta)^2}} \quad (6)$$

where $k^L = \frac{L_z}{L_y}$, $k^U = \frac{U_y}{U_z}$ are dimensionless fitting parameters. The ratio $\frac{k^U}{k^L} = \frac{J_c(90^\circ)}{J_c(0^\circ)}$ gives a rough estimate of the degree of anisotropy. The angle θ is measured from the plane of the sample.

The anisotropic pinning model was tested on low-temperature Nb–Ti superconductors and had not been applied to HTS tapes before this work.

## Experiment

**Experiment details.** The experiments presented in this paper consisted of measuring the current-voltage characteristics (CVC) of HTS tape samples in an external magnetic field of up to 1.5 T generated by a split magnet at a given field orientation in the configuration of the maximum Lorentz force. These measurements were conducted in a liquid nitrogen environment at ambient pressure, with a corresponding temperature of approximately 77.4 K. The CVCs were measured using the standard four-probe method. The sample size was as follows: length 80 mm, width 4 mm, thickness ~0.15 mm, distance between potential contacts of 10 mm between the voltages contacts located in the central region of the sample. The critical current was determined from the CVC in accordance with the conventional criterion of an electric field of 1 μV/cm. In instances where the objective was to determine the influence of the Lorentz force's orientation, the polarity of the transport current was reversed while the sample position remaining unchanged. Then the orientation angle of the external magnetic field $\theta$ was changed, and the CVC measurement was repeated. The angular dependence of the critical current $I_c(\theta)$ was established through the analysis of a series of volt-ampere characteristics measured in this way. The experimental apparatus was designed to facilitate a comprehensive rotation of 360°. The precision of the angle measurement was estimated to be no less than 0.5°.

**Samples.** The samples were cut from long-length scale HTS tapes (coated conductors), fabricated on the pilot line in the National Research Center 'Kurchatov Institute' [30]. All HTS tapes have a multilayer structure consisting of a stainless steel substrate tape (~100 μm), a buffer layer of yttrium-stabilized zirconium oxide YSZ (~2 μm), in which the texture is formed, a $CeO_2$ buffer layer (~0.2 μm), the HTS layer, a protective silver layer (~2 μm on each side of the tape), and a shunting copper layer (~20 μm on each side of the tape). The tapes were deferred in chemical composition and thickness of the HTS layer (Table 1). The substitution of rare earth elements within the REBCO composition can lead to a significant change in the pinning landscape [31, 32 thus rendering it pertinent to compare the angular dependences of the critical current of these HTS tapes.

Table1 – Description of samples

| Sample | Marking | Thickness of the HTS layer, μm | Critical Current in self-field, A | Field, T |
|---|---|---|---|---|
| YBCO | 282 | 1.40 | 84 | 0,1; 0,3; 0,5; 1; 1,5 |
| GdBCO | 365 | 1.05 | 148 | |
| SmBCO | 372-2 | 1.39 | 143 | |
| (Y+Sm)BCO | 393-2 | 1.60 | 101 | 1 |

**Angular dependences of the critical current.** Figures 3–5 show the measured critical current angular dependences for YBCO, GdBCO, and SmBCO samples in external fields of $\mu_0 H$ = 0.1, 0.3, 0.5, 1, and 1.5 T. The minimum value of the external field is chosen so as to clearly exceed the value of the self-field, which does not exceed 0.04 T. The angle θ is defined with respect to the plane of the tape. For all HTS tapes, significant anisotropy of $I_c$ is observed: the critical current is much higher when the field is directed in the plane of the tape (0, 180º) than when the field is oriented normal to the tape (90, 270º). In a field of 1 T, the ratio of the currents in these orientations is 6.9, 3.4, and 2.2 for YBCO, GdBCO, and SmBCO, respectively.

The GdBCO and SmBCO tapes exhibit peak asymmetry, which is clearly seen in polar coordinates by the absence of mirror symmetry between the upper and lower halves of the angular dependence (see insets in Figs. 3–5). Within the experimental accuracy, peak asymmetry in the angular dependence of the critical current is absent only in the YBCO sample. In addition, the SmBCO sample exhibits a significant dependence of the critical current on the current polarity, which was practically not observed for the YBCO and GdBCO samples. As illustrated in Fig. 5, for the current polarity represented by closed symbols, the peak corresponding to $\theta$ = 0º exhibits a greater magnitude than the peak at $\theta$ = 180º. Conversely, upon inverting the current direction (represented by open symbols), the ratio of peak amplitudes is inverted: the peak at θ = 0º is diminished relative to the peak at θ = 180º. The weaker the external field, the more pronounced this effect. At the external field strength of 1 T, the dependence of the critical current on the current polarity nearly vanishes.

Figure 6 shows the intricate influence of the rare earth element incorporated within the high-temperature superconducting (HTS) material on the angular dependence of the critical current. For the purposes of accurate comparison, the critical current is normalized to the current in the self-field. With a significant difference in the angular dependences of the YBCO and SmBCO samples, the angular dependence of the $(Y_{0.5}Sm_{0.5})$BCO sample does not exhibit an intermediary behavior, as one might anticipate, but rather a nearly complete absence of anisotropy: the minimum critical current is attained at approximately 45º (135º), rather than the anticipated 90º, and the critical current values at 0º and 90º are nearly equivalent.

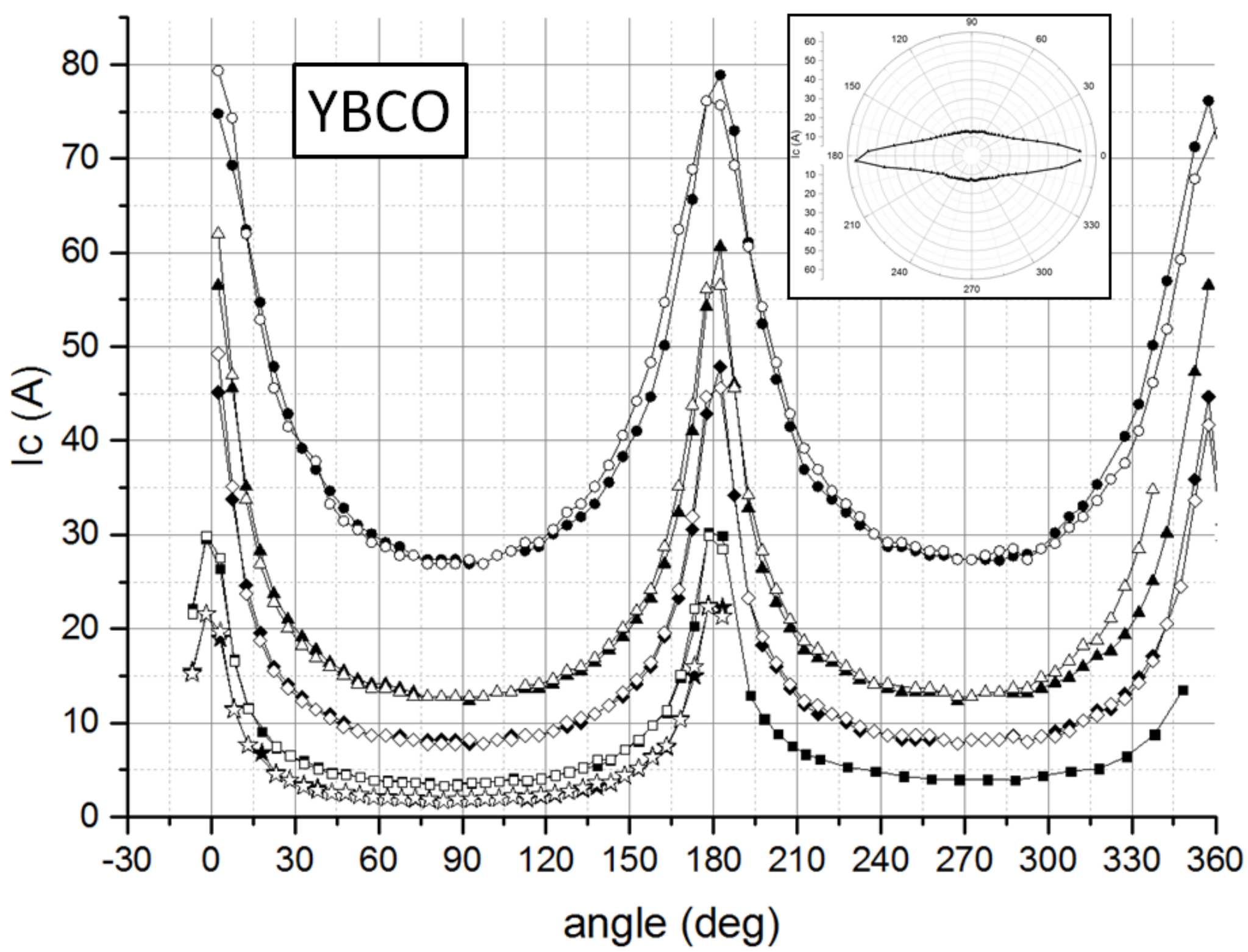

Fig. 3. Critical current angular dependences of of the YBCO sample in external fields (from top to bottom) of 0.1; 0.3; 0.5; 1 and 1.5 T. Closed and open symbols correspond to a change in the current polarity. In the inset, the angular dependence in a field of 0.3 T is reconstructed in polar coordinates to demonstrate the absence of peak asymmetry.

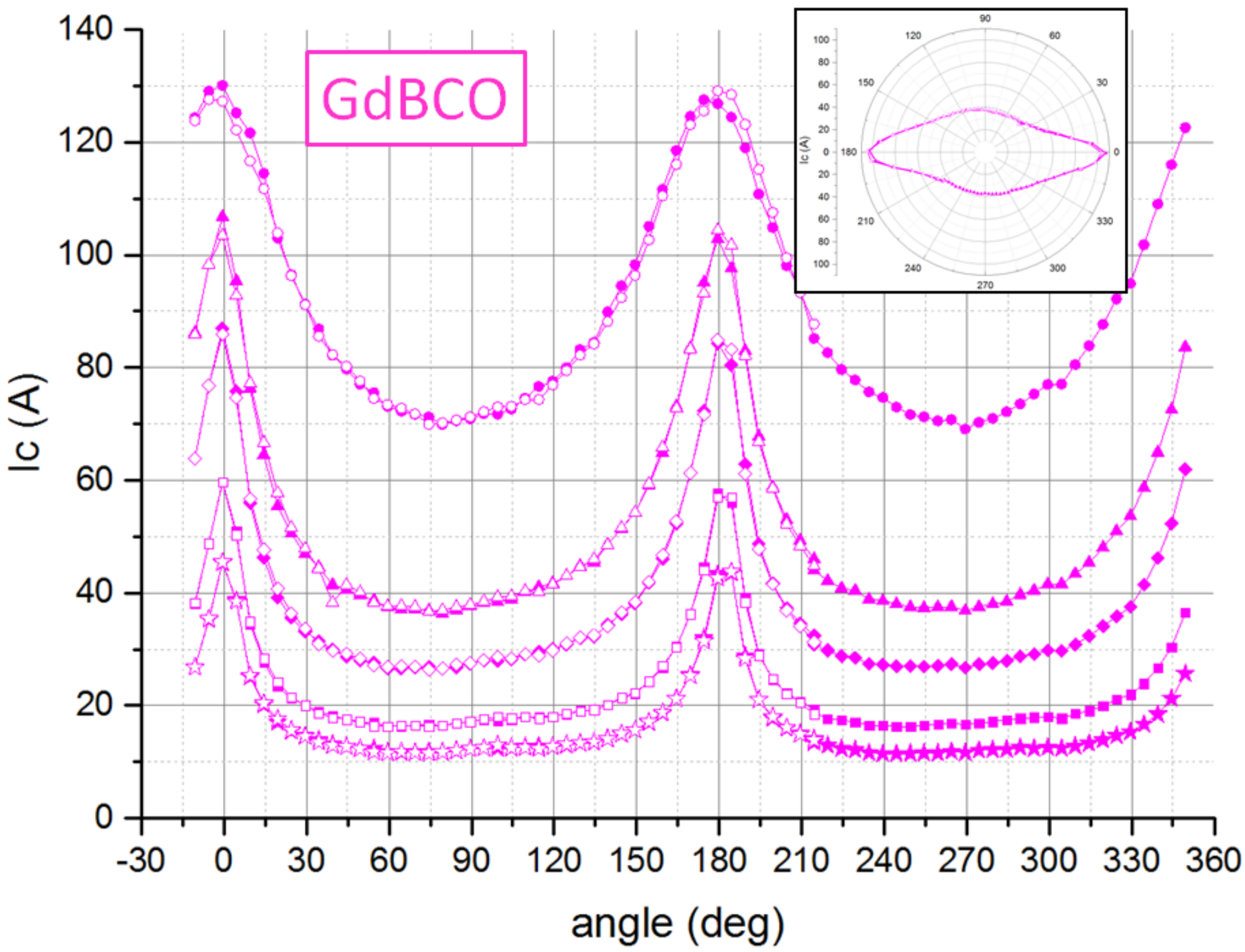

Fig. 4. Critical current angular dependences of the GdBCO sample in external fields (from top to bottom) of 0.1; 0.3; 0.5; 1 and 1.5 T. Closed and open symbols correspond to a change in the polarity of the current. The inset shows the angular dependence in a field of 0.3 T reconstructed in polar coordinates to demonstrate the asymmetry of the peaks.

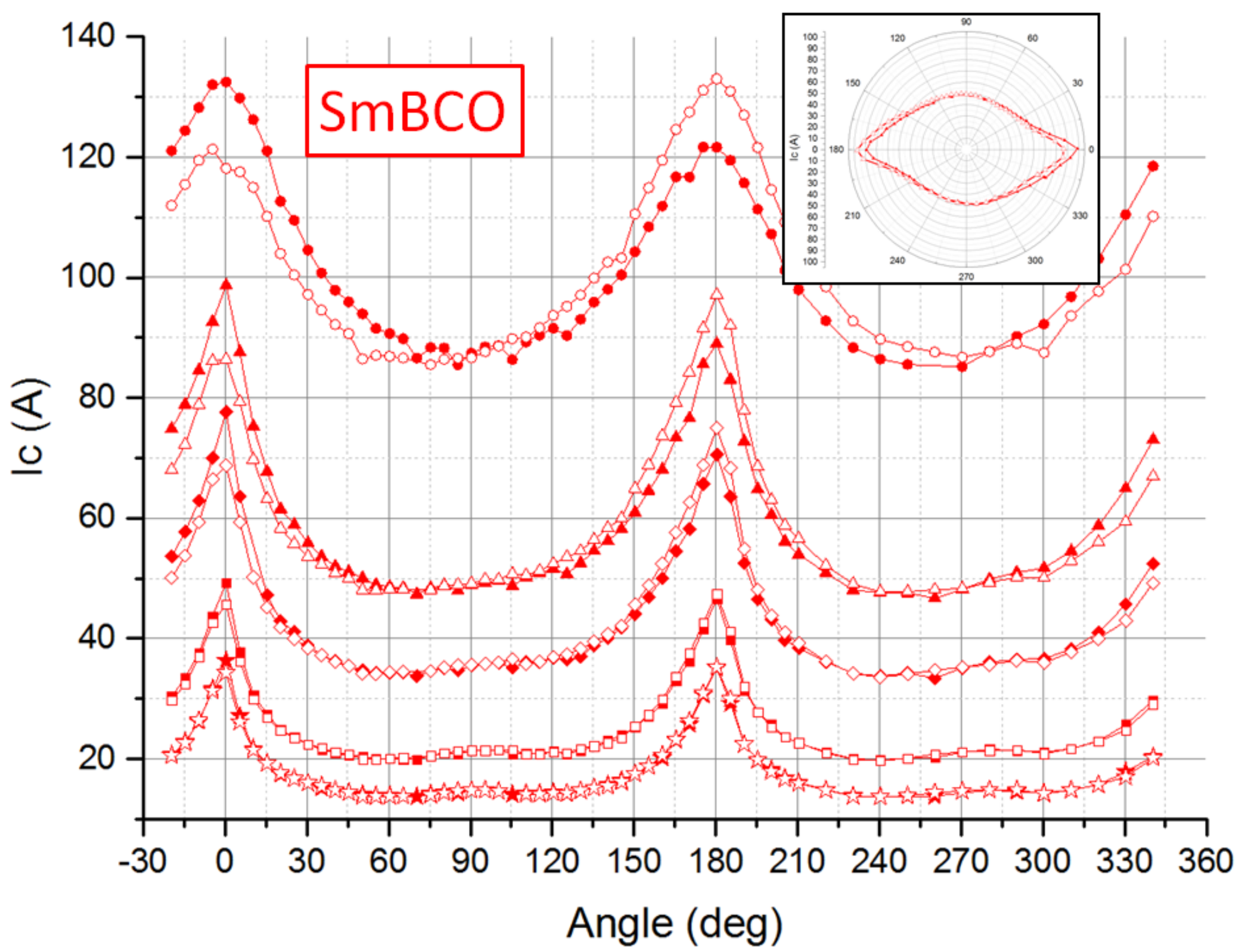

Fig. 5. Critical current angular dependences of the the SmBCO sample in external fields (from top to bottom) of 0.1; 0.3; 0.5; 1 and 1.5 T. Closed and open symbols correspond to a change in the polarity of the current. In the inset, the angular dependence in a field of 0.3 T is reconstructed in polar coordinates to demonstrate the asymmetry of the peaks.

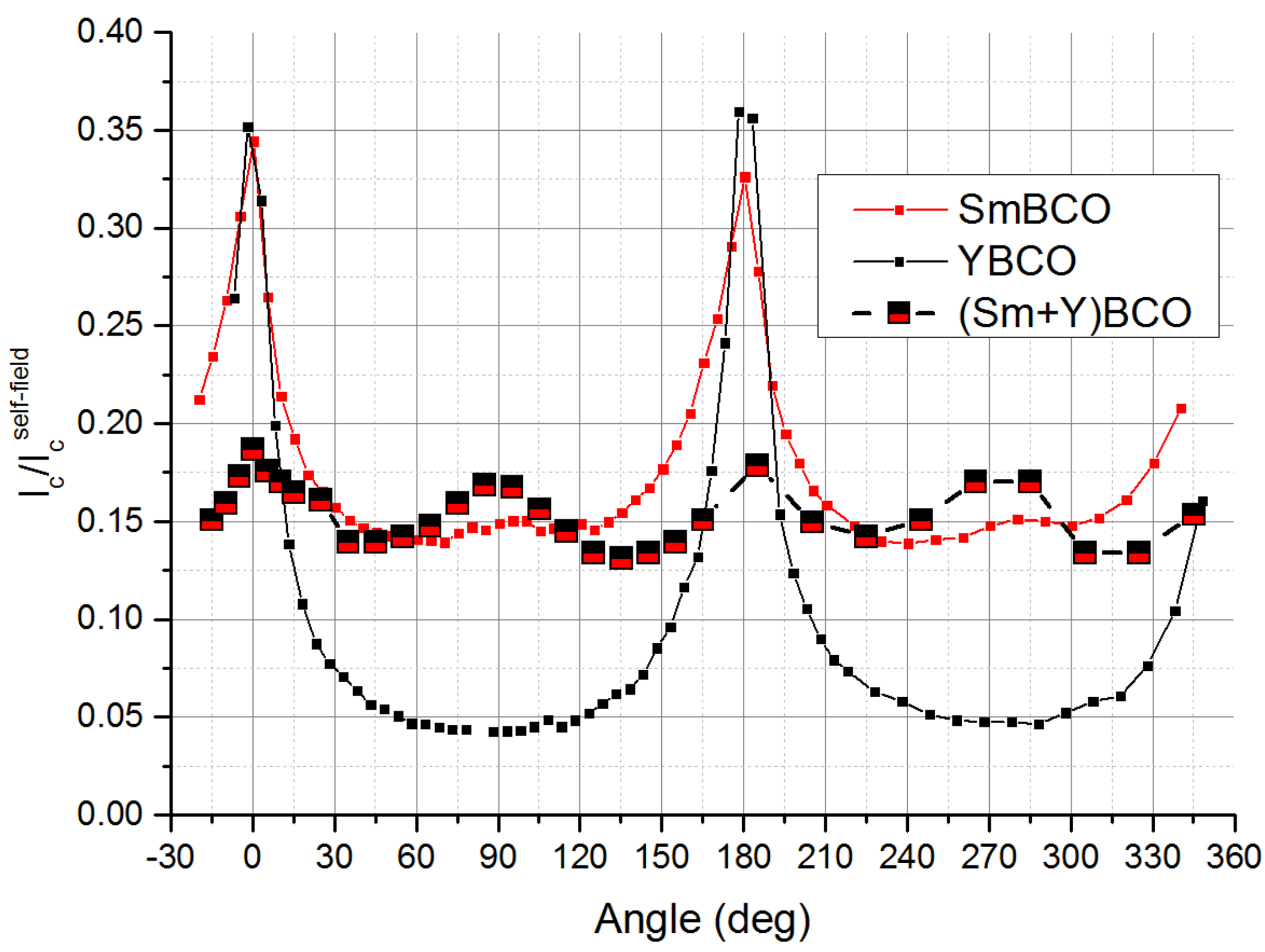

Fig. 6. The reduced critical current angular dependences of of YBCO SmBCO and (Sm+Y)BCO samples in a field of 1 T.

## Comparison of the experimental observations with theoretical models

**Scaling model.** The experimental angular dependences $I_c(\theta)$ manifest qualitative deviations from the functional dependence (1). Such deviations are not entirely unforeseen, as the anisotropy mechanism associated with the effective mass of charge carriers does not imply a description of all phenomena, including the emergence of additional peaks (Fig. 6, $(Y_{0.5}Sm_{0.5})$BCO tape), peak asymmetry (Figs. 4 and 5, GdBCO and SmBCO tapes) and the dependence of the critical current on the current polarity (Fig. 5, SmBCO tape). Nevertheless, even in instances where the aforementioned features are absent (Fig. 4, YBCO tape), the predictions of this model are not fully realized, as the coefficient γ depends on the magnitude of the external field and deviates significantly from the predicted limits 5-7 (Fig. 7). As a consequence, when rearranging in the coordinates $I_c$—$H/f(\theta)$, no convergence to the scaling curve is observed (Fig. 8).

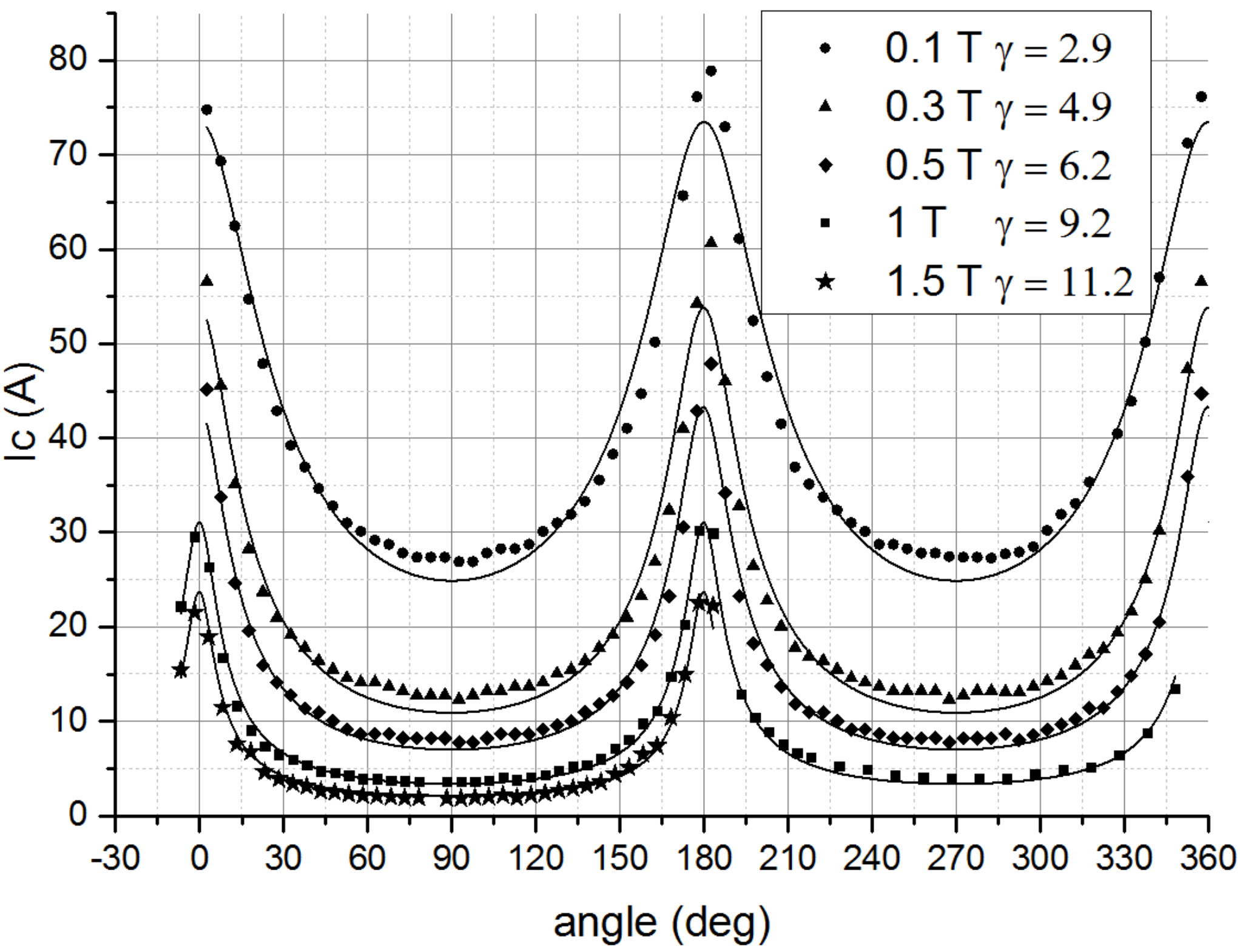

Fig. 7. The dots show the experimental angular dependences of the YBCO sample. The solid lines are the fit using formula (1). The magnitude of the external field and the fitting parameter γ from formula (1) are indicated in the legend.

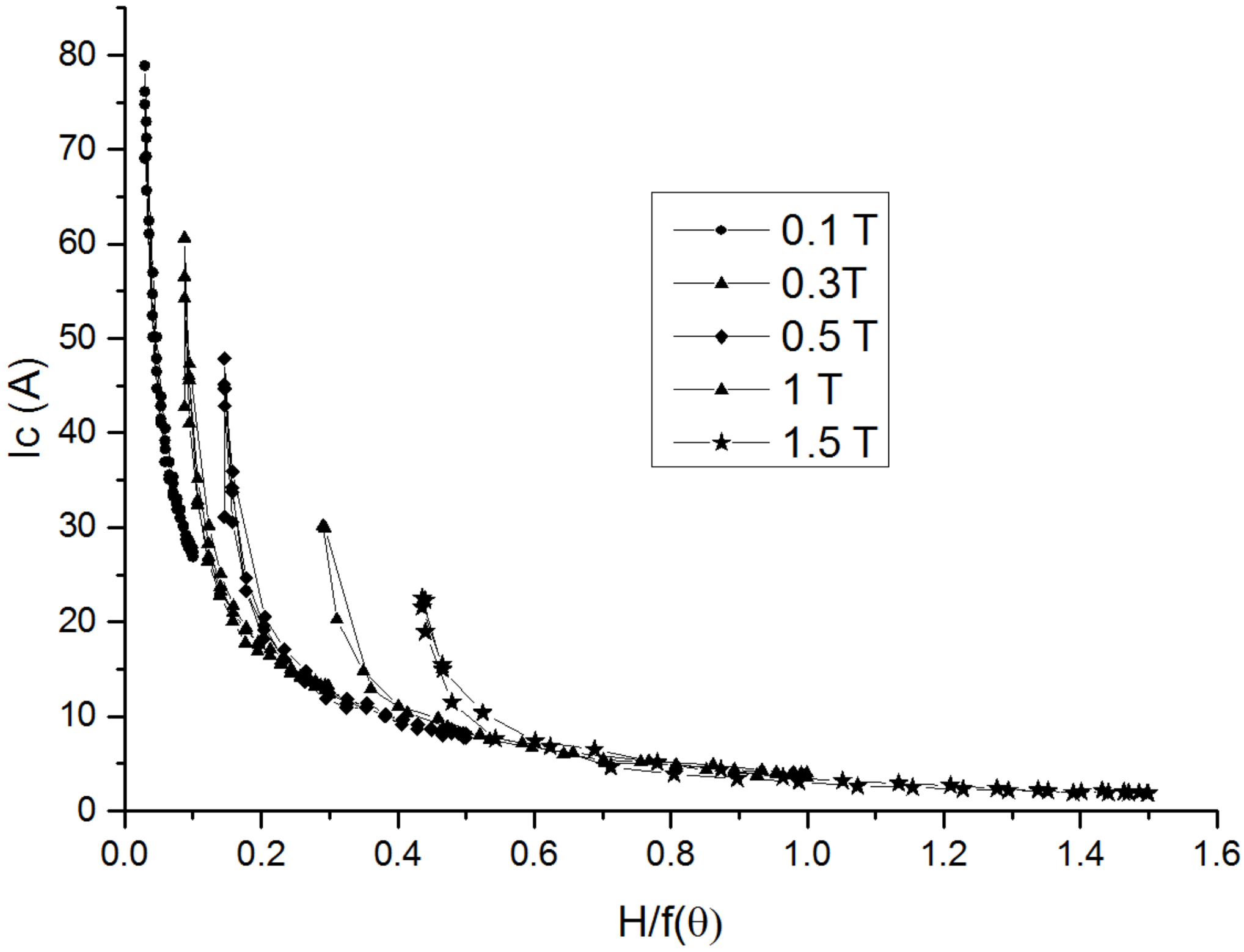

Fig. 8. The data in Fig. 7 reconstructed in the coordinates Ic – H/f(θ) in accordance with the scaling concept. The average parameter $\gamma$=6.7 was used for the construction.

Table 2 presents the values of the fitting coefficient $\gamma$ for the analyzed tapes along with the quality of the fit, which is characterized by the determination coefficient $R$, adjusted for the quantity of fitting parameters in the model. The proximity of the $R$-value to unity indicates the efficacy of the model in describing the experimental data. For the YBCO tape, the value of $R$ exceeds 0.95 in all fields, which, considering that the model employs only a single fitting parameter, represents an exceptionally favorable outcome. In contrast, for the other tapes, the alignment of the experimental results with the functional dependence (1) is considerably poorer: for GdBCO, the parameter R falls below 0.86, while for SmBCO, it drops below 0.78. For the $(Y_{0.5}Sm_{0.5})$BCO tape, the parameter R approaches 0, as evidenced by the sharp deviation of the angular dependence from the model predictions depicted in Fig. 6.

Table 2 – The value of the fitting parameter $\gamma$ and some statistical characteristics of the approximating curves characterizing the quality of the fit..

| Sample | Field, T | $\gamma$ | Coefficient of determination $R$ |
|---|---|---|---|
| YBCO | 0.1 | 2.9 | 0.9746 |
| | 0.3 | 4.9 | 0.9597 |
| | 0.5 | 6.2 | 0.96414 |
| | 1 | 9.2 | 0.98901 |
| | 1.5 | 11.2 | 0.98945 |
| GdBCO | 0.1 | 1.8 | 0.97843 |
| | 0.3 | 2.9 | 0.92914 |
| | 0.5 | 3.4 | 0.89989 |
|  | 1 | 3.9 | 0.85552 |
|  | 1.5 | 4.4 | 0.85238 |
| SmBCO | 0.1 | 1.5 | 0.90945 |
|  | 0.3 | 1.9 | 0.85526 |
|  | 0.5 | 2.1 | 0.8193 |
|  | 1 | 2.2 | 0.77028 |
|  | 1.5 | 2.5 | 0.77374 |
| (Y+Sm)BCO | 1 | - | - |

**Vortex path model**. Figures 9 and 10 show examples of processing angular dependencies within the vortex path model framework. Due to the large number of adjustable parameters (three parameters per peak plus one "setpoint" parameter), the model describes the experiment very well. The peak asymmetry is described by two subsystems of pinning centers with close positions of the averages (Fig. 9), the additional peak at the orientation normal to the tape is described by an additional Lorentzian or Gaussian (Fig. 10).

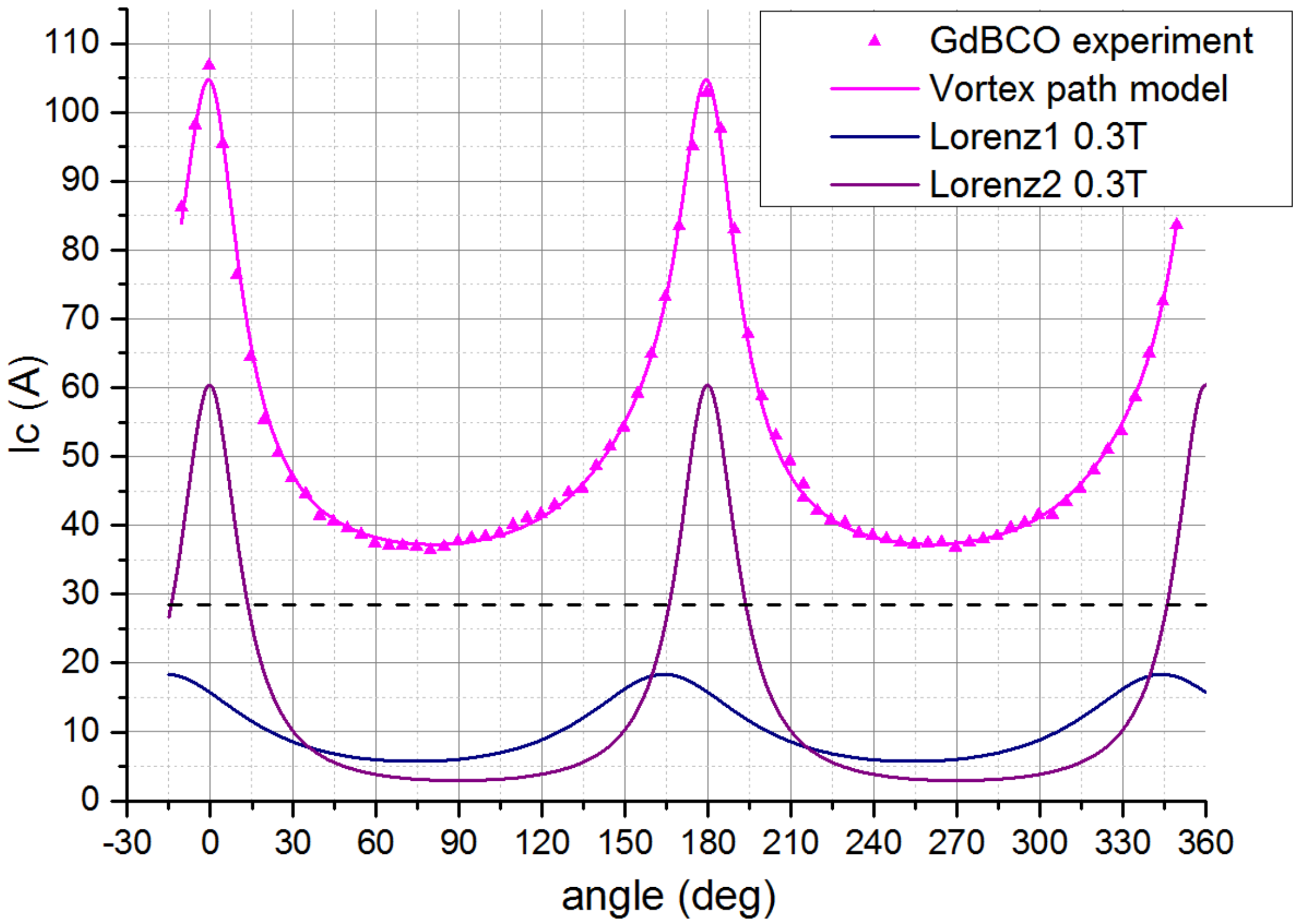

Fig. 8. An example of approximation of the experimental angular dependence with a pronounced peak asymmetry for the GdBCO sample at 0.3 T – closed symbols. The asymmetry is taken into account by two distributions with close peak positions, shown in blue and violet. Their linear sum with a constant (dashed line) gives the approximation – a line passing through the experimental points.

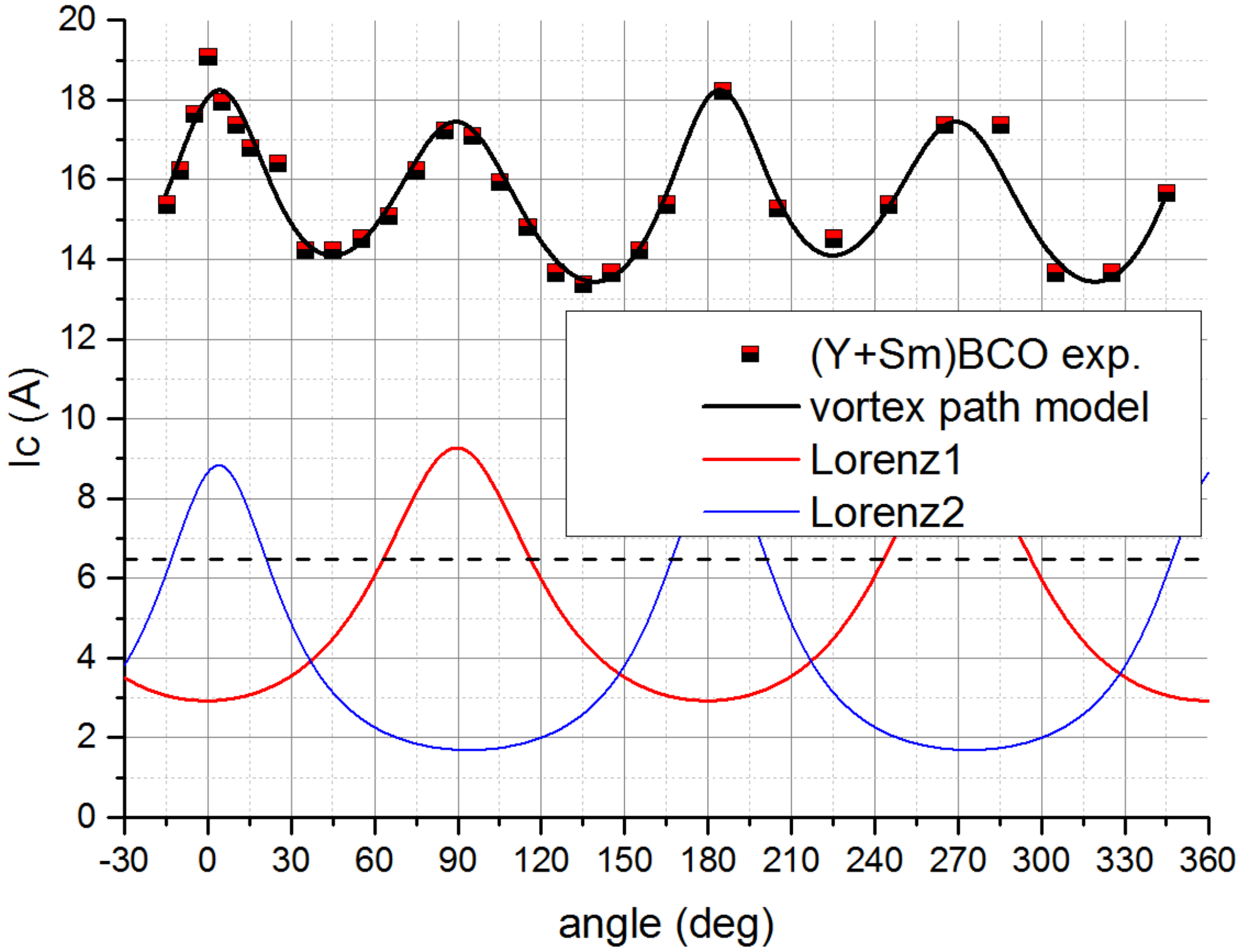

Fig. 9. – An example of approximation of the experimental angular dependence (point) having a peak not only near θ=0°, 180° (field in the plane of the tape), but also an additional peak at θ=90°, 270° when the field is oriented along the normal to the tape for the (Y+Sm)BCO sample in a field of 1 T. The main peak is approximated by the Lorentzian shown by the blue line, the additional peak is taken into account by the additional Lorentzian shown by the red line.

The incorporation of the required number of Lorentzians or Gaussians allows describing almost all the features observed in the angular dependences. The exception is the dependence of the peak height on the polarity of the current or, what is the same, the different heights of the peaks at orientations of 0 and 180º. This is due to the fact that the periodicity of the approximation functions (2) and (3) is 180º.

Table 3 shows the values of the fitting coefficients for the studied samples and the quality of the fit. The YBCO sample, which has no features in the angular dependence, is well described by one Lorentzian with the determination coefficient *R* of no less than 0.96. In order to account for the asymmetry present in the GdBCO and SmBCO samples, it becomes necessary to introduce an additional Lorentzian. At the same time, in a field of 0.1 T, the additional Lorentzian yields no significant enhancement in the quality of the fit. The dependence of the critical current value on the current polarity for the SmBCO sample leads to a lower determination coefficient compared to GdBCO: 0.944 against 0.995 at 0.1 T. For all approximation models employed, the determination coefficient in this paradigm consistently exceeds 0.9.

Table 3 – Values of the fitting parameters of the vortex path model and some statistical characteristics of the approximating curves characterizing the quality of the fit.

| Sample | Field, T | $I_0$ | Lorenz 1 | Lorenz 2 | Coefficient of determination *R* |
|---|---|---|---|---|---|
| YBCO | 0,1 | 20,95 | $I_1$=59,82<br>$\theta_1$=0<br>$\Gamma_1$=2,87 | | 0.99037 |
| | 0,3 | 11,0 | $I_1$=32,15<br>$\theta_1$=0,08<br>$\Gamma_1$=4,33 | | 0,97574 |
| | 0,5 | 6,87 | $I_1$=24,38<br>$\theta_1$=-0,23<br>$\Gamma_1$=4,73 | | 0,96947 |
| | 1 | 3,53 | $I_1$=13,95<br>$\theta_1$=-0,20<br>$\Gamma_1$=6,08 | | 0,98703 |
| | 1,5 | 1,69 | $I_1$=10,44<br>$\theta_1$=-0,33<br>$\Gamma_1$=6,30 | | 0,98742 |
| GdBCO | 0,1 | 53,07 | $I_1$=115,15<br>$\theta_1$=-2,89<br>$\Gamma_1$=2,059 | | 0,99533 |
| | 0,3 | 28,48 | $I_1$=47,74<br>$\theta_1$=-0,32<br>$\Gamma_1$=4,54 | $I_2$=32,17<br>$\theta_2$=163,0<br>$\Gamma_2$=1,79 | 0,99669 |
| | 0,5 | 22,47 | $I_1$=30,47<br>$\theta_1$=0,15<br>$\Gamma_1$=5,54 | $I_2$=18,20<br>$\theta_2$=164,01<br>$\Gamma_2$=2,17 | 0,99482 |
| | 1 | 14,95 | $I_1$=16,72<br>$\theta_1$=0,93<br>$\Gamma_1$=7,25 | $I_2$=9,41<br>$\theta_2$=168,76<br>$\Gamma_2$=2,90 | 0,99255 |
| | 1,5 | 10,68 | $I_1$=10,63<br>$\theta_1$=1,44<br>$\Gamma_1$=8,46 | $I_2$=8,22<br>$\theta_2$=173,17<br>$\Gamma_2$=3,2 | 0,99141 |
| SmBCO | 0,1 | 73,32 | $I_1$=85,0<br>$\theta_1$=-2,39<br>$\Gamma_1$=1,98 | | 0,94369 |
| | 0,3 | 42,05 | $I_1$=13,42<br>$\theta_1$=0,86<br>$\Gamma_1$=6,32 | $I_2$=38,41<br>$\theta_2$=170,73<br>$\Gamma_2$=2,26 | 0,97852 |
| | 0,5 | 31,08 | $I_1$=10,11<br>$\theta_1$=0,62<br>$\Gamma_1$=7,81 | $I_2$=24,85<br>$\theta_2$=170,53<br>$\Gamma_2$=2,64 | 0,98752 |
| | 1 | 18,90 | $I_1$=5,88<br>$\theta_1$=0,45 | $I_2$=13,15<br>$\theta_2$=172,90 | 0,99226 |
|  |  |  | $\Gamma_1$=9,26 | $\Gamma_2$=3,24 |  |
|  | 1,5 | 12,83 | $I_1$=4,39<br>$\theta_1$=0,14<br>$\Gamma_1$=9,85 | $I_2$=10,08<br>$\theta_2$=174,6<br>$\Gamma_2$=3,11 | 0,99431 |
| (Y+Sm)BCO | 1 | 6,49 | $I_1$=16,34<br>$\theta_1$=-0,42<br>$\Gamma_1$=0,56 | $I_2$=12,13<br>$\theta_2$=93,99<br>$\Gamma_2$=0,44 | 0,92049 |

**Anisotropic pinning model.** Table 4 presents the values of the fitting coefficients of the anisotropic pinning model (7) for the analyzed tapes. Figure 11 shows the approximation of the experimental angular dependences of the YBCO. With increasing field, a tendency for the $k^L$ parameter to decrease from 1.44 (at 0.3 T) to 1 (at 1.5 T) is observed. When $k^L = 1$, the angular dependence (7) coincides with function (1) of the scaling model. Within the framework of the anisotropic pinning model, this specific condition corresponds to a scenario in which the dimensional ellipsoid is transformed into a spherical shape, allowing for two energetically equivalent states of the vortex ensemble to exist at the same spatial distance in any given direction. This scenario can be realized either in relatively strong magnetic fields, when the distance between two energetically equivalent positions is determined by the intervortex distance, i.e. the magnitude of the external magnetic field [28], or within a sparse arrangement of pinning centers or in systems characterized by weak collective pinning, even under relatively low magnetic fields – a phenomenon frequently observed in single-crystal samples. This condition partially elucidates the rationale behind the ascendance of the scaling model, since the initial studies of HTS materials were focused mainly on single crystals.

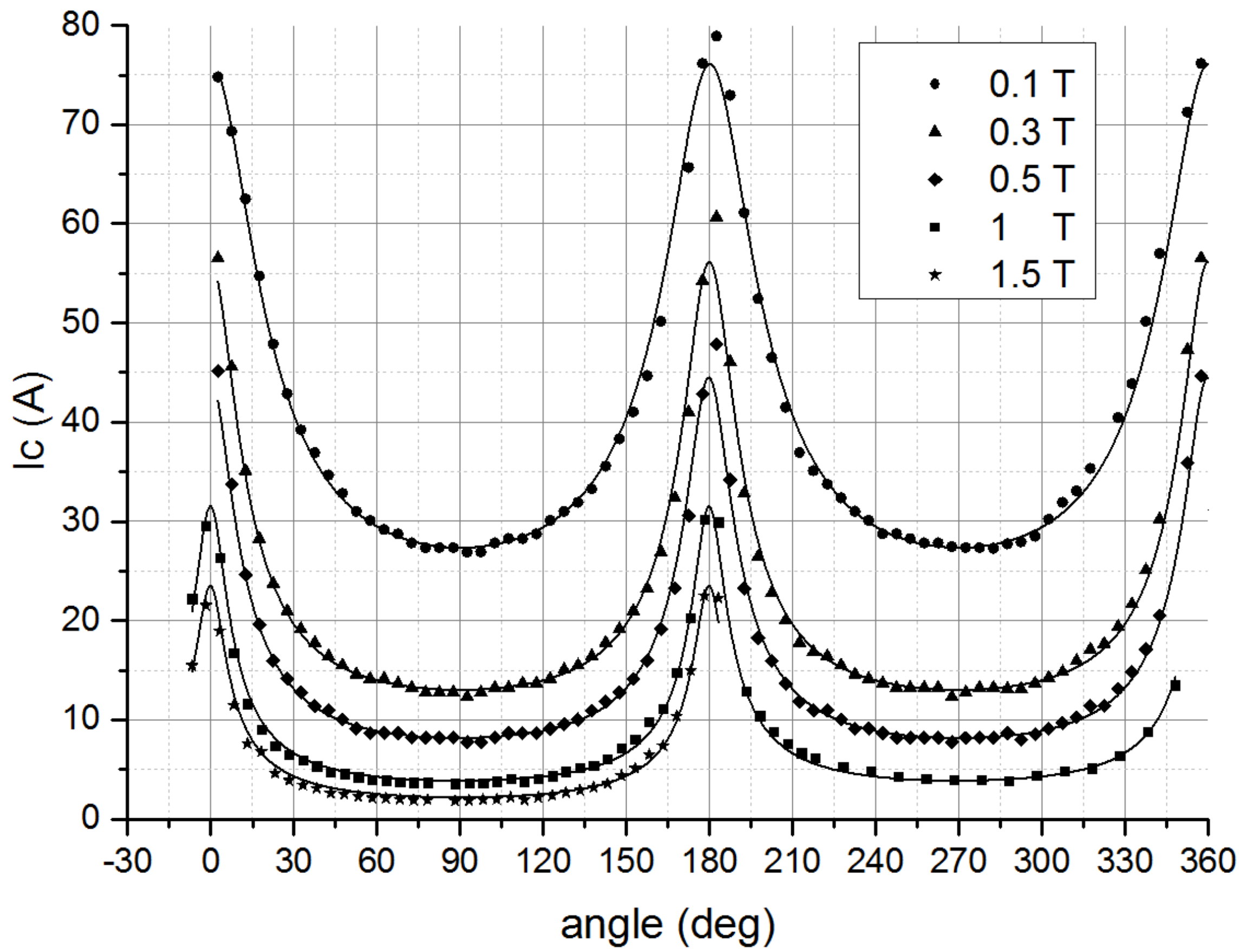

Fig. 11 Example of approximation of experimental angular dependences of critical current for YBCO sample within the framework of the anisotropic pinning model.

The determination coefficient $R$ for the YBCO does not fall below 0.97, indicating a commendable outcome (Table 4).

The anisotropic pinning model, as delineated in Section 2, fails to adequately account for the existence of asymmetric peaks within the angular dependence of $I_c(\theta)$. To address this limitation, an additional fitting parameter, denoted as $\theta_0$, has been introduced, which characterizes the deviation of the principal axes of the energy ellipsoid from those of the dimensional ellipsoid (as illustrated in Fig. 12). Various authors have repeatedly noted a correlation between the presence of peak asymmetry and the deviation of the crystallographic *c* axis from the normal to the tape plana by several degrees [33]. It is reasonable to assume the energy ellipsoid is influenced by the spatial orientation of the pinning center system, namely, by defects that develop in conjunction with the crystallographic orientation, while the dimensional ellipsoid pertains to the geometry of the tape. Indeed, the demagnetizing factor changes from a zero value when the field is oriented in the plane of the tape to a maximum value when the field is oriented along the normal, which, according to many authors, plays a significant role in the manifestation of peak asymmetry [34]. Thus, we obtain the following approximating dependence:

$$I_c(\theta) = I_c^0 \sqrt{\frac{[k^L \cos\theta]^2 + [\sin\theta]^2}{[k^U \cos(\theta - \theta_0)]^2 + [\sin(\theta - \theta_0)]^2}} \quad (7)$$

Figure 13 shows the approximation for the GdBCO within the framework of the anisotropic pinning model, taking into consideration the introduced parameter $\theta_0$. The determination coefficient surpasses 0.99 for all magnetic fields (Table 4). The APM model incorporates the anisotropy of the critical current not solely in relation to the direction of the magnetic field but also concerning the orientation of the Lorentz force. In fact, such an effective description of the peak asymmetry is a direct result of acknowledging this dual nature of anisotropy. A more pronounced dependence on the Lorentz force direction is observed in the SmBCO tape, where not only is peak asymmetry evident, but also a significant difference in the critical current value depending on the direction of its flow (polarity). This new feature can be integrated into the APM model through an additional modification—substituting the dimensional ellipsoid with an egg-shaped spatial figure. The section of this figure is an ellipse, the semi-axes of which exhibit differing magnitudes (Fig. 12). The approximation of the critical current angular dependence for the SmBCO tape, considering these modifications, is presented in Fig. 14, with the determination coefficient for all magnetic fields being no less than 0.98 (Table 4).

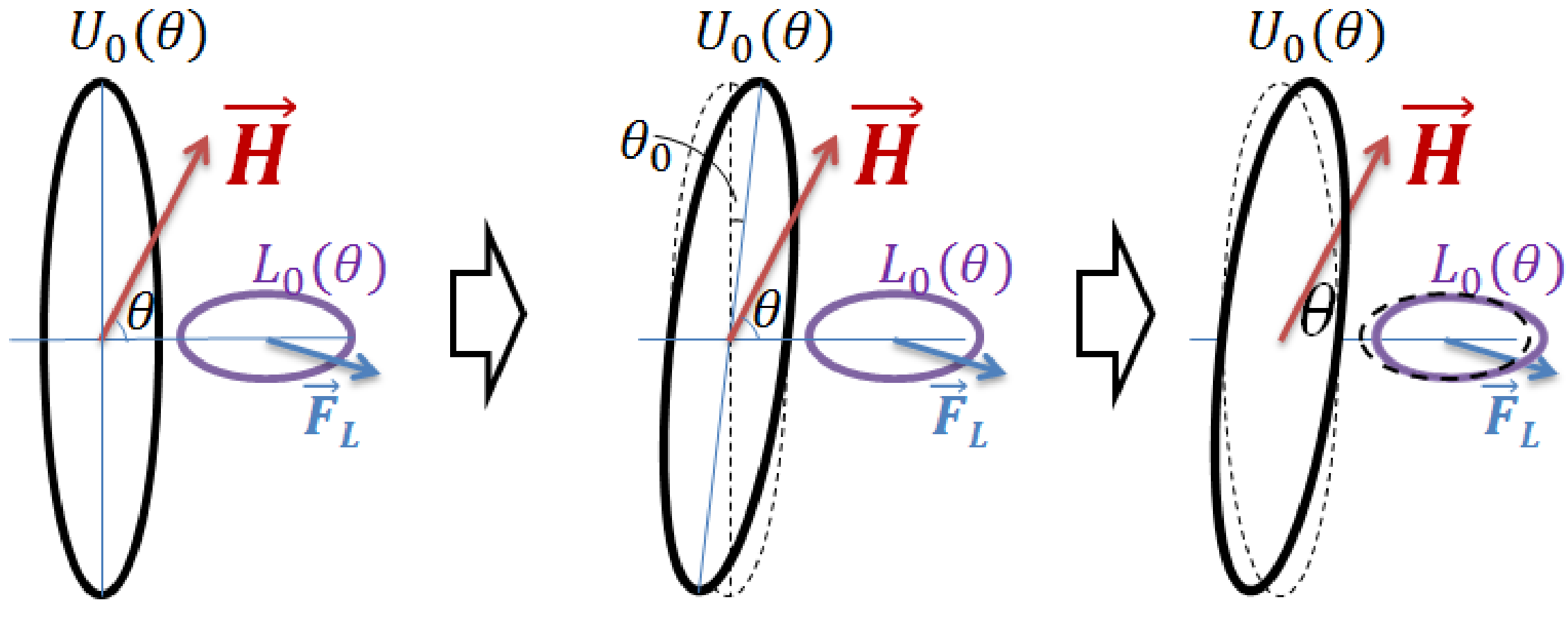

Fig. 12. Schema of the proposed modifications of the anisotropic pinning model. On the left is a schematic representation of the cross-section of the energy and size ellipsoids for the considered geometry of the maximum Lorentz force and the definition of the angle $\theta$. In the center is a diagram of the introduction of the parameter $\theta_0$ as the angle of deviation of the main axes of the energy and size ellipses. On the right is a modification diagram to take into account the dependence of the critical current on the direction of the Lorentz force - the right and left semi-axes of the size ellipsoid do not coincide.

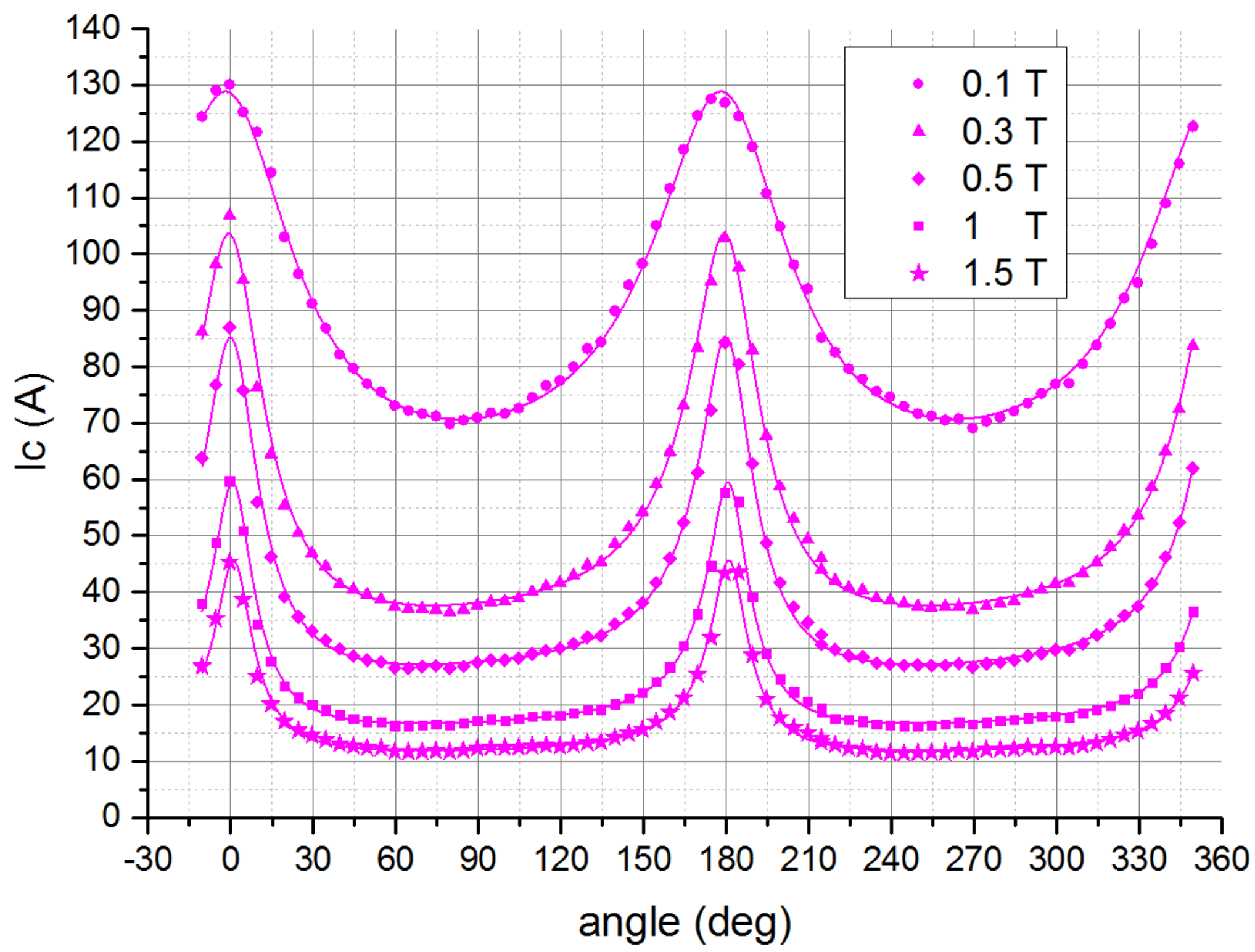

Fig. 13. An example of approximation of experimental angular dependences of the critical current for a GdBCO sample within the framework of the anisotropic pinning model.

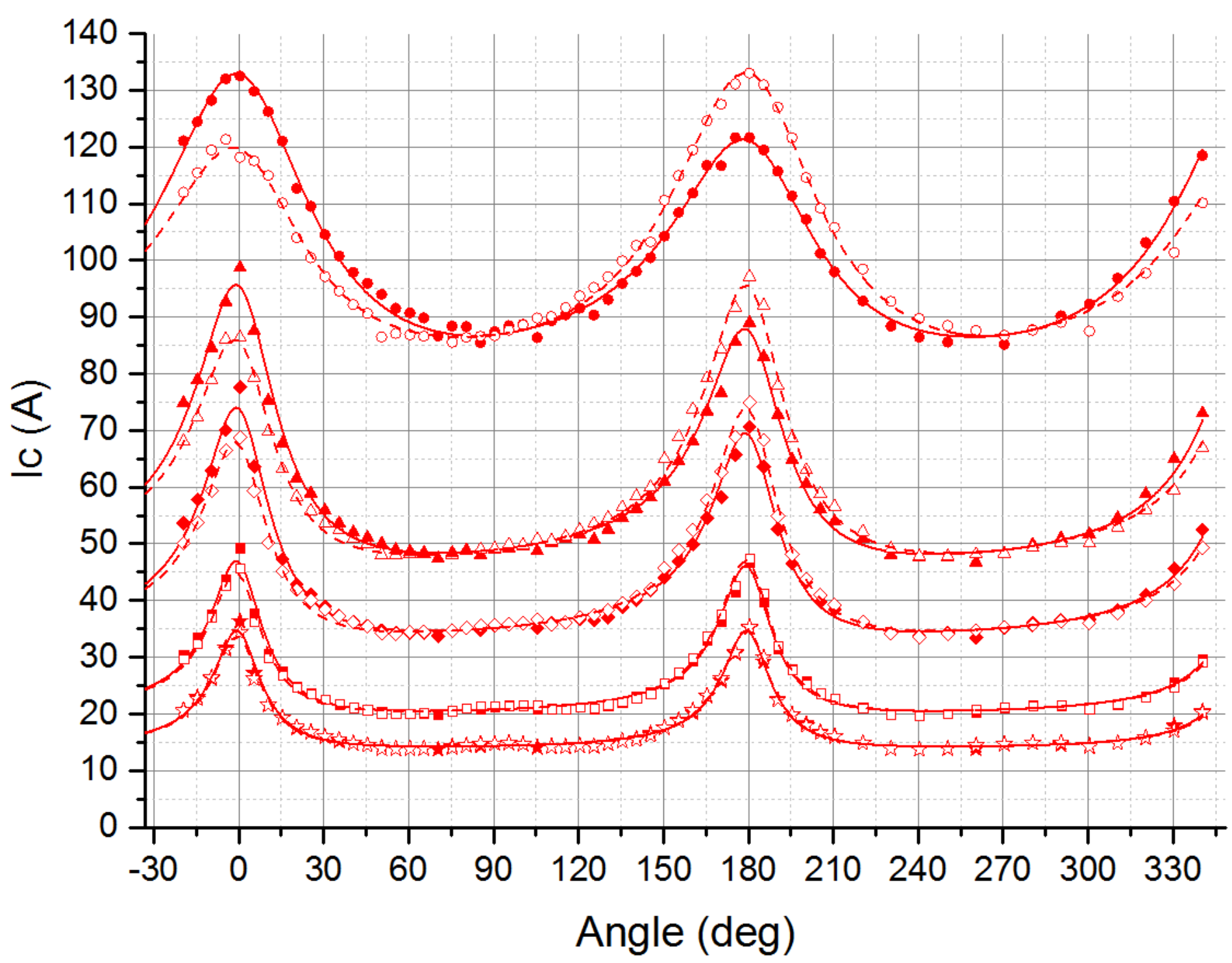

Fig. 14. An example of approximation of experimental critical current angular dependences for SmBCO sample within the framework of the anisotropic pinning model.

In order to accurately characterize the additional peaks observed in the angular dependence of the critical current in the $(Y_{0.5}Sm_{0.5})$BCO, particularly when the external magnetic field is oriented perpendicularly to the tape, it is imperative to implement more substantial modifications than those previously delineated. One potential alteration involves the abandonment of the ellipsoidal representation of energy and/or dimensional entities in favor of a more intricate geometric configuration. Indeed, the energy and dimensional bodies of the Nb-Ti and HTS-2 tapes are not necessary required to have the same shape. It is plausible that the successful modeling of the experimental data pertaining to the YBCO, GdBCO, and SmBCO samples within this approximation is largely coincidental. However, considering that the requisite modifications may be introduced in various ways, further investigations are warranted to ascertain the most rational strategy for amending the model.

Without introducing the appropriate modifications to the model, specifically the alterations necessary to account for the additional peaks, the coefficient of determination experiences a reduction to 0.5.

Table 4 – Values of the fitting parameters of the anisotropic pinning model and the statistical parameter of the approximating curves characterizing the quality of the fit.

| Sample | Field, T | fitting parameters | Coefficient of determination $R$ |
|---|---|---|---|
| YBCO | 0.1 | $I_c^0$=76.08; $k^L$ =1.33; $k^U$=3.71 | 0.991 |
| | 0.3 | $I_c^0$=56.15; $k^L$ =1.44; $k^U$=6.20 | 0.977 |
| | 0.5 | $I_c^0$=44.50; $k^L$ =1.31; $k^U$=7.12 | 0.971 |
| | 1 | $I_c^0$=31.54; $k^L$ =1.21; $k^U$=9.86 | 0.991 |
| | 1.5 | $I_c^0$=23.50; $k^L$ =1; $k^U$=10.71 | 0.989 |
| GdBCO | 0.1 | $I_c^0$=128.28; $\theta_0$=5.58; $k^L$ =1.33; $k^U$=2.4 | 0.997 |
| | 0.3 | $I_c^0$=101.74; $\theta_0$=7.02; $k^L$ =1.8; $k^U$=4.77 | 0.996 |
| | 0.5 | $I_c^0$=83.36; $\theta_0$=6.62; $k^L$ =2.02; $k^U$=6.07 | 0.995 |
| | 1 | $I_c^0$=58.29; $\theta_0$=5.26; $k^L$ =2.34; $k^U$=7.97 | 0.993 |
| | 1.5 | $I_c^0$=44.95; $\theta_0$=4.45; $k^L$ =2.38; $k^U$=8.84 | 0.991 |
| SmBCO | 0.1 | $I_c^0$=86.98; $\theta_0$=-3.8; $k_1^L$ =1.44; $k_2^L$ =1.59; $k^U$=2.19 | 0.993 |
| | 0.3 | $I_c^0$=49.07; $\theta_0$=-4.7; $k_1^L$ =2.06; $k_2^L$ =2.26; $k^U$=3.97 | 0.990 |
| | 0.5 | $I_c^0$=35.26; $\theta_0$=-4.5; $k_1^L$ =2.35; $k_2^L$ =2.54; $k^U$=4.84 | 0.989 |
| | 1 | $I_c^0$=20.82; $\theta_0$=-3.27; $k_1^L$ =2.75; $k_2^L$ =2.85; $k^U$=6.15 | 0.987 |
| | 1.5 | $I_c^0$=14.35; $\theta_0$=-2.5; $k_1^L$ =2.82; $k_2^L$ =2.87; $k^U$=6.78 | 0.987 |
| (Y+Sm)BCO | 1 | $I_c^0$=18.22; $\theta_0$=-1.63; $k^L$ =5.59; $k^U$=6.72 | 0.525 |

## Conclusion

In this paper, we present experimental critical current angular dependences for four REBCO tapes with different rare earth elements. Drawing upon the acquired data, alongside findings from other researchers, we can formulate the following empirical generalizations.

I. The substitution of the rare earth element within the REBCO significantly changes the angular dependence of the critical current. These changes cannot be reduced to the dependence on the ionic radius size. The angular dependence of the critical current is predominantly governed by the anisotropy inherent in the pinning center configuration, rather than the anisotropic characteristics of the REBCO crystallographic unit cell.

II. The critical current demonstrates a dualistic nature of anisotropy: it is influenced not solely by the orientation of the magnetic field, but also by the direction of the Lorentz force, which is dictated by both current and magnetic field directions

The paper provides a brief overview of existing models used to analyze the angular dependences. The models under consideration have been evaluated against experimental data.

The scaling method is the most prevalent technique for analyzing the angular dependences of the critical current. This approach is quite elegant since it relies on only one fitting parameter and gives fairly strong predictions for the behavior of the angular dependence of the critical current, thereby minimizing the necessity for subjective determinations during data processing. However, the application of this method to the description of practical superconductors with strong pinning has neither a convincing theoretical justification nor agreement with experimental data.

The vortex path model presents an alternative approach that has gained popularity recently. This model is based on a more realistic assumption that pinning plays a predominant role in the anisotropy of the critical current. It is postulated that the observed peaks can be analyzed by analogy with signal processing methods commonly used in spectroscopy or X-ray diffraction.The model does not impose strict limitations on the number of fitting parameters, which allows one to achieve near-perfect agreement between the experiment and the approximating curve by introducing an increasing number of them. However, it is crucial to recognize that a satisfactory description of the data does not invariably ensure a correct interpretation. For example, the asymmetry of the peaks on $I_c(\theta)$ in this model is rationalized by the existence of at least two subsystems of pinning centers that possess similar yet non-coinciding angular orientations, which seems farfetched. The lack of a clear physical interpretation of the fitting parameters complicates the establishment of a linkage with structural characteristics. Furthermore, this model fails to provide an adequate representation of the observed dependence of the critical current value on the direction of the Lorentz force.

Finally, the third and most promising model, in our opinion, is the anisotropic pinning model. This framework was developed based on the notion of dualism in the anisotropy of critical current: anisotropy concerning both the orientation of the magnetic field and that of the Lorentz force. Within this model, the asymmetry of the peaks is construed as a manifestation of this dualism, which is more plausible than the interpretation of the vortex path model. With the value of the fitting parameter $k^L = 1$, the predictions of this model coincide with the functional dependence of the scaling model. The possibility of a connection between the fitting parameters and the structure of the material is traced. When modeling angular dependencies characterized by a single peak within the range of –90 º to +90º, this model yields the highest determination coefficient, which indicates a better approximation. However, at the present stage of development, the existing model remains inadequate for describing additional peaks that may appear in the experimental angular dependencies when the orientation of the external magnetic field deviates from the plane of the superconducting tape. This limitation underscores the necessity for further refinement of the model.

This work was carried out within the framework of the state assignment of the National Research Center "Kurchatov Institute".